\documentclass{icrc2009}

\usepackage{graphicx}   
\usepackage[font=footnotesize,caption=false]{subfig} 
\usepackage{fixltx2e}
\usepackage{url}

\newcommand{\shorttitle}[1]%
{\markboth{Proceedings of the 31\MakeLowercase{$^{st}$} ICRC, {\L}\'{o}d\'{z} 2009}{#1} }
\newcommand{\etal}{\MakeLowercase{\textit{et al. }}} 

\begin{document}
\title{Sensor development and calibration for acoustic\\
  neutrino detection in ice}

\author{\IEEEauthorblockN{Timo Karg\IEEEauthorrefmark{1}, Martin
    Bissok\IEEEauthorrefmark{2}, Karim Laihem\IEEEauthorrefmark{2},
    Benjamin Semburg\IEEEauthorrefmark{1}, and Delia
    Tosi\IEEEauthorrefmark{3}\\
    for the IceCube collaboration\IEEEauthorrefmark{4}}\\
  \IEEEauthorblockA{\IEEEauthorrefmark{1}Dept. of Physics, University
    of Wuppertal, D-42119 Wuppertal, Germany}
  \IEEEauthorblockA{\IEEEauthorrefmark{2}III Physikalisches Institut,
    RWTH Aachen University, D-52056 Aachen, Germany}
  \IEEEauthorblockA{\IEEEauthorrefmark{3}DESY, D-15735 Zeuthen,
    Germany}
  \IEEEauthorblockA{\IEEEauthorrefmark{4}See
    http://www.icecube.wisc.edu/collaboration/authorlists/2009/4.html}}

\shorttitle{Karg \etal Acoustic sensor calibration}
\maketitle

\begin{abstract}
  A promising approach to measure the expected low flux of cosmic
  neutrinos at the highest energies (E $>$ 1\,EeV) is acoustic
  detection. There are different in-situ test installations worldwide
  in water and ice to measure the acoustic properties of the medium
  with regard to the feasibility of acoustic neutrino detection. The
  parameters of interest include attenuation length, sound speed
  profile, background noise level and transient backgrounds. The South
  Pole Acoustic Test Setup (SPATS) has been deployed in the upper
  500\,m of drill holes for the IceCube neutrino observatory at the
  geographic South Pole. In-situ calibration of sensors under the
  combined influence of low temperature, high ambient pressure, and
  ice-sensor acoustic coupling is difficult. We discuss laboratory
  calibrations in water and ice. Two new laboratory facilities, the
  Aachen Acoustic Laboratory (AAL) and the Wuppertal Water Tank Test
  Facility, have been set up. They offer large volumes of bubble free
  ice (3\,m$^3$) and water (11\,m$^3$) for the development, testing,
  and calibration of acoustic sensors. Furthermore, these facilities
  allow for verification of the thermoacoustic model of sound
  generation through energy deposition in the ice by a pulsed
  laser. Results from laboratory measurements to disentangle the
  effects of the different environmental influences and to test the
  thermoacoustic model are presented.
\end{abstract}


\section{Introduction}
\label{sec:introduction}

The detection and spectroscopy of extra-terrestrial ultra high energy
neutrinos would allow us to gain new insights in the fields of
astroparticle and particle physics. Apart from the possibility to
study particle acceleration in cosmic sources, the measurement of the
guaranteed flux of cosmogenic neutrinos \cite{Berezinsky:1969} opens a
new window to study cosmic source evolution and particle physics at
unprecedented center of mass energies. However, the fluxes predicted
for those neutrinos are very low \cite{Engel:2001}, so detectors with
large target masses are required for their detection. One possibility
to instrument volumes of ice of the order of $100\,\mathrm{km}^3$ with
a reasonable number of sensor channels is to detect the acoustic
signal emitted from the particle cascade at a neutrino interaction
vertex \cite{Askaryan:1957}.

To study the properties of Antarctic ice relevant for acoustic
neutrino detection the South Pole Acoustic Test Setup (SPATS)
\cite{Vandenbroucke:2009} has been frozen into the upper part of
IceCube \cite{IceCube} boreholes. SPATS consists of four vertical
strings reaching a depth of 500\,m below the surface. The horizontal
distances between strings cover the range from 125\,m to 543\,m. Each
string is instrumented with seven acoustic sensors and seven
transmitters. The ice parameters to be measured are the sound speed
profile, the acoustic attenuation length, the background noise level,
and transient noise events in the frequency range from 1\,kHz to
100\,kHz.

For the design of a large scale acoustic neutrino detector it is
crucial to fully understand the {\em in-situ} response of the sensors
as well as the thermoacoustic sound generation mechanism.

\section{Sensor calibration}
\label{sec:calibration}

To study the acoustic properties of the Antarctic ice, like the
absolute background noise level, and to deduce the arrival direction
and energy of a neutrino in a future acoustic neutrino telescope it is
essential to measure the sensitivity and directionality of the sensors
used, i.e.~the output voltage as function of the incident pressure,
and its variation with the arrival direction of the incident acoustic
wave relative to the sensor. These measurements can be carried out
relatively easily in the laboratory in liquid water. The two
calibration methods most commonly used are

\begin{itemize}
\item the comparison method, where an acoustic signal sent by a
  transmitter (with negligible angular variation) is simultaneously
  recorded at equal distance with a pre-calibrated receiver and the
  sensor to be calibrated. A comparison of the signal amplitudes in
  the two receivers allows for the derivation of the desired
  sensitivity from the sensitivity of the pre-calibrated sensor.
\item the reciprocity method, which makes use of the electroacoustic
  reciprocity principle to determine the sensitivity of an acoustic
  receiver without having to use a pre-calibrated receiver (see
  e.g.~\cite{Urick:1983}).
\end{itemize}

All SPATS sensors have been calibrated in $0 \, ^\circ$C water with
the comparison method \cite{Fischer:2006}. However, both calibration
methods are not suitable for in-situ calibration of sensors in South
Pole ice. There are no pre-calibrated sensors for ice available, and
reciprocity calibration requires large setups which are not feasible
for deployment in IceCube boreholes. Further, directionality studies
require a change of relative positioning between emitter and receiver
which is difficult to achieve in a frozen-in setup.

It is not clear how results obtained in the laboratory in liquid water
can be transferred to an in-situ situation where the sensors are
frozen into Antarctic ice. We are studying the influence of the
following three environmental parameters on the sensitivity
separately: low temperature, increased ambient pressure, and different
acoustic coupling to the sensor. We will assume that sensitivity
variations due to these factors obtained separately can be combined to
a total sensitivity change for frozen in transmitters. This assumption
can then be checked further using the two different sensor types
deployed with the SPATS setup. Apart from the standard SPATS sensors
with steel housing two HADES type sensors \cite{Semburg:2009} have
been deployed with the fourth SPATS string. These contain a
piezoceramic sensor cast in resin and are believed to have different
systematics.

\subsection{Low temperatures}
\label{sec:temperature}

\begin{figure}[!t]
  \centering
  \includegraphics[width=8cm]{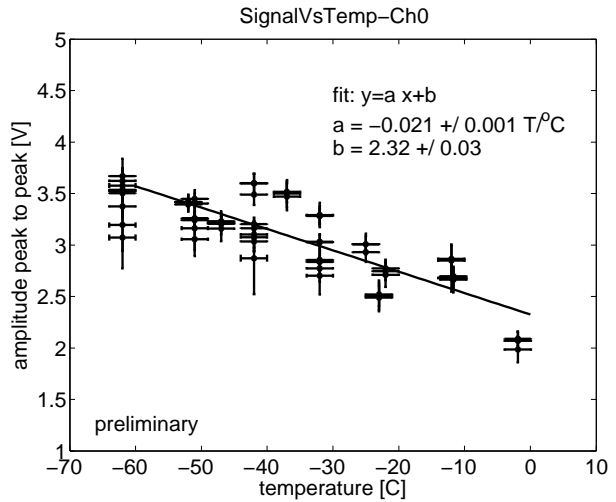}
  \caption{Measured peak-to-peak amplitude of one SPATS sensor channel
    in air at different temperatures and linear fit to the data.}
  \label{fig:zeuthen_V_vs_T}
\end{figure}

The ice temperature in the upper few hundred meters of South Pole ice
is $-50 \, ^\circ$C \cite{Price:2002}. It is not feasible to produce
laboratory ice at this temperature in a large enough volume to carry
out calibration studies. We study the dependence of the sensitivity on
temperature in air. A signal sent by an emitter is recorded with a
sensor at different temperatures.  To prevent changes in the
emissivity of the transmitter, the transmitter is kept at constant
temperature outside the freezer, and only the sensor is cooled down.
The recorded peak-to-peak amplitude is used as a measure of
sensitivity. First results indicate a linear increase of sensitivity
with decreasing temperature (cf.~Fig.~\ref{fig:zeuthen_V_vs_T}). The
sensitivity of a SPATS sensor is increased by a factor of $1.5 \pm
0.2$ when the temperature is lowered from $0 \, ^\circ$C to $-50 \,
^\circ$C (averaged over all three sensor channels).

\subsection{Static pressure}
\label{sec:pressure}

Acoustic sensors in deep polar ice are exposed to increased ambient
static pressure. During deployment this pressure is exerted by the
water column in the borehole (max.~50\,bar at 500\,m depth). During
re-freezing it increases since the hole freezes from the top,
developing a confined water volume. The pressure is believed to
decrease slowly as strain in the hole ice equilibrates to the bulk ice
volume. The final static pressure on the sensor is unknown.

A $40.5$\,cm inner diameter pressure vessel is available at Uppsala
university that allows for studies of sensor sensitivity as function
of ambient pressure. Static pressures between 0 and 800\,bar can be
reached. In this study the pressure is increased up to
100\,bar. Acoustic emitters for calibration purposes can be placed
inside the vessel or, free of pressure, outside of it. Cable feeds
allow one to operate up to two sensors or transmitters inside the
vessel.  A sensor is placed in the center of the water filled
vessel. The transmitter is coupled from the outside to the vessel. The
recorded peak-to-peak amplitude is used as a measure of sensitivity
while the pressure is increased. The sensor sensitivity is measured by
transmitting single cycle gated sine wave signals with different
central frequencies from 5\,kHz to 100\,kHz.

\begin{figure}[!t]
  \centering
  \includegraphics[width=8cm]{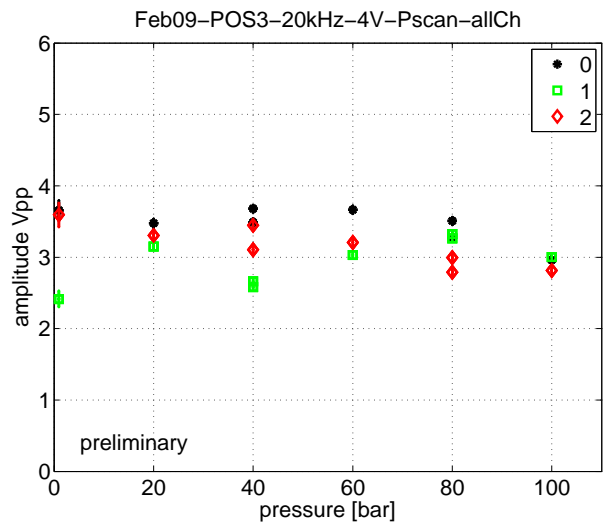}
  \caption{Measured peak-to-peak amplitude of a SPATS sensor excited
    by a transmitter coupled from the outside to the pressure
    vessel. All three channels of the sensor are shown.}
  \label{fig:uppsala_V_vs_p}
\end{figure}

Figure \ref{fig:uppsala_V_vs_p} shows the received signal amplitudes
for the three sensor channels of a SPATS sensor as a function of
ambient pressure. No systematic variation of the sensitivity with
ambient pressure is observed. Combining all available data we conclude
that the variation of sensitivity with static pressure is less than
30\% for pressures below 100\,bar.

\subsection{Sensor-ice acoustic coupling}
\label{sec:coupling}

The acoustic coupling, i.e.~the fractions of signal energy transmitted
and reflected at the interface of medium and sensor, differs
significantly between water and ice. It can be determined using the
characteristic acoustic impedance of the medium and sensor, which is
the product of density and sound velocity and is equivalent to the
index of refraction in optics. Due to the different sound speeds the
characteristic acoustic impedance of ice is about $2.5$ times higher
than in water.

Its influence will be studied in the Aachen Acoustic Laboratory
(Sec.~\ref{sec:facilities}), where it will be possible to carry out
reciprocal sensor calibrations in both water and ice, and also to use
laser induced thermoacoustic signals as a calibrated sound source.

\section{New laboratory facilities}
\label{sec:facilities}

Two new laboratories have been made available to the IceCube Acoustic
Neutrino Detection working group for signal generation studies and
sensor development and calibration.

\paragraph{Wuppertal Water Tank Test Facility}

For rapid prototyping of sensors and calibration studies in water, the
Wuppertal Water Tank Test Facility offers a cylindrical water tank
with a diameter of $2.5$\,m and depth of $2.3$\,m ($11 \,
\mathrm{m}^3$). The tank is built up from stacked concrete rings and
has a walkable platform on top. It is equipped with a positioning
system for sensors and transmitters and a 16-channel PC based DAQ
system (National Instruments USB-6251 BNC).

The size of the water volume allows for the clean separation of
emitted acoustic signals and their reflections from the walls and
surface. This makes it possible to install triangular reciprocity
calibration setups with side lengths of up to 1\,m. Further,
installations to measure the polar and azimuthal sensitivity of
sensors are possible.

\paragraph{Aachen Acoustic Laboratory}

\begin{figure}[!t]
  \centering
  \includegraphics[width=7.5cm]{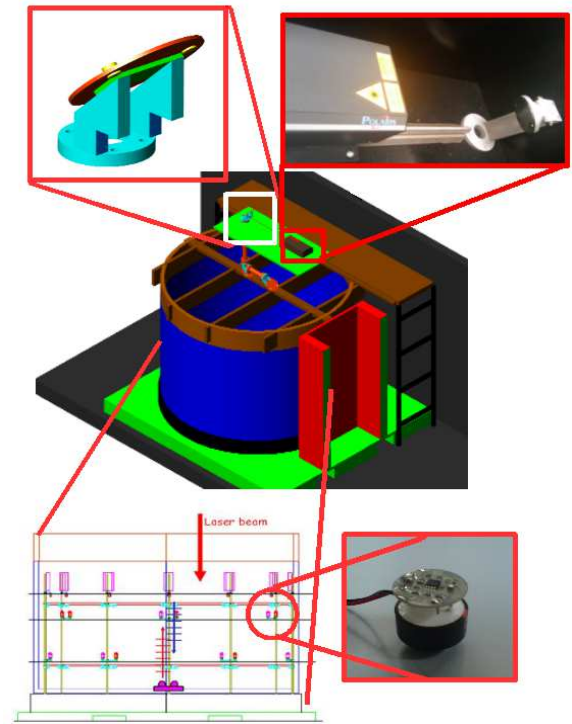}
  \caption{Overview of the AAL setup with zoom on the mirror holder
    (top), the sensor positioning system (bottom, left) and a sensor
    (bottom, right).}
  \label{fig:aal_setup}
\end{figure}

The Aachen Acoustic Laboratory is dedicated to the study of
thermoacoustic sound generation in ice. A schematic overview of the
setup can be seen in Fig.~\ref{fig:aal_setup}. The main part is a
commercial cooling container ($6 \times 2.5 \times 2.5 \,
\mathrm{m}^3$), which can reach temperatures down to $-25 \,
^\circ$C. An IceTop tank, an open cylindrical plastic tank with a
diameter of 190\,cm and a height of 100\,cm \cite{Gaisser:2005}, is
located inside the container. The IceTop tank has a freeze control
unit by means of which the production of bubble-free ice is
possible. The freeze control unit mainly consists of a cylindrical
semipermeable membrane at the bottom of the container, which is
connected to a vacuum reservoir and a pressure regulation system. The
membrane allows for degassing of the water. A total volume of $\approx
3 \, \mathrm{m}^3$ of bubble-free ice can be produced. A full freezing
cycle takes approximately sixty days with the freezing going from top
to bottom.

On top of the container, a Nd:YAG Laser is installed in a light-tight
box with an interlock connected to the laser control unit. The laser
has a pulse repetition rate of up to 20\,Hz and a peak energy per
pulse of 55\,mJ at 1064\,nm, 30\,mJ at 532\,nm, and 7\,mJ at 355\,nm
wavelength. The laser beam is guided into the container and deposited
in variable positions on the ice surface by a set of mirrors with
coatings for the above mentioned frequencies. The optical feed-through
consists of a tilted quartz window to avoid damage of the laser cavity
by reflected laser light. For the detection of thermoacoustic signals,
18 sensors are mounted on a sensor positioning system. The positioning
system has three levels, on each level 6 sensors are placed in a
hexagonal geometry. Along with the sensors, 18 sound emitters are
deployed for calibration and test purposes. The sensors will be
calibrated reciprocally. The positioning system will also include a
reciprocal calibration setup for HADES sensors and the ability to
install a SPATS sensor for calibration purposes. In addition, two
temperature sensors are deployed at each level.  The acoustic sensors
are pre-amplified low-cost piezo based ultrasound sensors, usually
used for distance measurement. The sensors show a strong variation of
signal strength with incident angle. This directionality has to be
studied but is rather useful for the suppression of reflected signals.
The sensors are read out continuously by a LabVIEW-based DAQ framework
with a NI PCIe-6259M DAQ card. The framework includes a temperature
and acoustic noise monitoring system.

\section{Studies of thermoacoustic signal generation}
\label{sec:signal_generation}

A detailed understanding of thermoacoustic sound generation in ice is
crucial for designing an acoustic extension to the IceCube
detector. The dependence of the signal strength on the deposited
energy as well as on the distance to the sensor is of great interest.
Also the pulse shape and the frequency content have to be studied
systematically with respect to various cascade parameters. The spatial
distribution of the acoustic signal has to be investigated, i.e.~the
acoustic disk and its dependence on the spatial and temporal energy
deposition distribution. In addition, the AAL setup will be able to
study the thermoacoustic effect in a wide temperature range from $20
\, ^\circ$C to $-25 \,^\circ$C and possible differences of the effect
in ice and water.

In the Aachen Acoustic Laboratory, the thermoacoustic signal is
generated by a Nd:YAG laser. A laser-induced thermoacoustic signal
differs from a signal produced in a hadronic cascade. While a
cascade's energy deposition profile can be described by a
Gaisser-Hillas function, the laser intensity drops of
exponentially. Also the lateral profile of a cascade follows a NKG
function, where, assuming a TEM$_{00}$ mode in the far field region,
the typical laser-beam profile is Gaussian.  Knowing this, a
recalculation of the signal properties from a laser-induced pulse to a
cascade-generated pulse is possible.

The frequency content of the signal is expected to vary with the beam
diameter, while a too short penetration depth will result in an
acoustic point source rather than a line source. The absorption
coefficient of light in water or ice varies strongly with
wavelength. The first wavelength of the laser (1064\,nm) is absorbed
after few centimeters, while the second harmonic (532\,nm) has an
absorption length of $\approx 60$\,m. The third harmonic at 355\,nm
with an absorption length of $\approx 1$\,m is expected to be the most
suitable wavelength to emulate a hadronic cascade with a typical
length of 10\,m and a diameter of 10\,cm. The diameter of the heated
ice volume has to be controlled by optics inside the container that
will widen the beam.

\begin{figure}[!t]
  \centering
  \includegraphics[width=8cm]{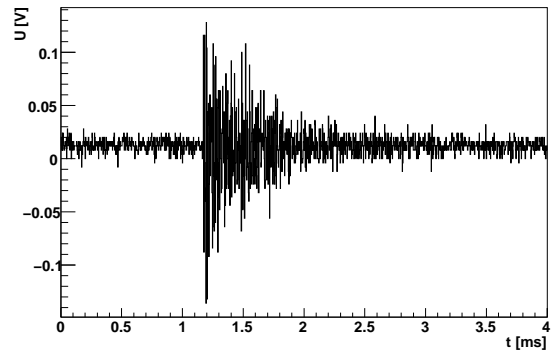}
  \caption{Thermoacoustic pulse in ice, generated with a laser at
    1064\,nm wavelength and a beam diameter of $\approx 1 \,
    \mathrm{\mu m}$.}
  \label{fig:thermoacoustic_pulse}
\end{figure}

With an array of 18 sensors, the Aachen Acoustic Laboratory allows the
study of the spatial distribution of the generated sound field, as
well as the frequency content with varying beam parameters.  The first
thermoacoustic signal has been generated and detected in a test setup
with preliminary sensor electronics in order to determine a reasonable
gain for the pre-amplifier while avoiding saturation.  A small volume
of bubble-free ice has been produced, containing a sensor and an
emitter. Laser pulses (wavelength 1064\,nm, 55\,mJ per pulse) have
been shot at the ice block. A zoom on the first waveform is presented
in Fig.~\ref{fig:thermoacoustic_pulse}. The distance between laser
spot and sensor is approximately $15\,\mathrm{cm}$ and the sensor gain
factor is 22. A Fourier transform of the pulse implies a pulse central
frequency of $\approx 100 \, \mathrm{kHz}$, which is expected for such
a small beam diameter. In order to see the expected bipolar pulse,
further studies have to be performed to determine the transfer
function of the sensors.

\section{Conclusions}
\label{sec:conclusion}

Detailed understanding of the thermoacoustic sound generation
mechanism and the response of acoustic sensors in Antarctic ice is
necessary to design an acoustic extension for the IceCube neutrino
telescope. While in-situ calibrations in deep South Pole ice are
inherently difficult, different environmental influences can be
studied separately in the laboratory. No change in sensor response
with increasing ambient pressure was found; a linear increase in
sensitivity with decreasing temperature was observed. Intense pulsed
laser beams can be used to generate thermoacoustic signals in ice
which can also be used as an in-ice calibration source.

\section*{Acknowledgments}

We are grateful for the support of the U.S.~National Science
Foundation and the hospitality of the NSF Amundsen-Scott South Pole
Station.

This work was supported by the German Ministry for Education and
Research.

\end{document}